# Crack opening and closure detection through coupled DCPD and non-continuous DIC method – application to LCF tests

Théotime ASSELIN[1,2*], Olivier ANCELET[2], Guillaume BENOIT[1], Florence HAMON[1], Valéry VALLE[1], Gilbert HÉNAFF[1]

[1]: Institut Pprime, Département Physique et Mécanique de Matériaux, École Nationale Supérieure de Mécanique et d'Aérotechnique, Téléport 2, 1 avenue Clément Ader, 86961 Futuroscope Chasseneuil, France.

[2]: Framatome, 1 place Jean Millier, 92048 Paris La Défense, France.

**Corresponding author:** Théotime ASSELIN - Mail: theotime.asselin@framatome.com - Adress: 1 place Jean Millier, 92048 Paris La Défense, France

**Keywords:**

Fatigue, Crack closure, Digital Image Correlation, Low-alloy steel, Direct Current Potential Drop.

**Abbreviations:**

DIC: Digital Image Correlation

DCPD: Direct Current Potential Drop

ACPD: Alternating Current Potential Drop

CMOD: Crack Mouth Opening Displacement

EDM: Electrical Discharge Machining

H-DIC: Heaviside Digital Image Correlation

LCF: Low Cycle Fatigue

**Nomenclature:**

| | |
|---|---|
| $\sigma_{YS}$ | 0,2% plastic strain offset yield strength |
| $\sigma_u$ | Tensile strength |
| $A$ | Tensile elongation |
| $\Delta\varepsilon_t$ | Strain variation |
| $\dot{\varepsilon}$ | Strain rate |
| $c$ | Total crack width |
| $a_{tot}$ | Total crack depth |
| $a_{def}$ | Initial defect depth |
| $V$ | Potential drop |
| $V_0$ | Reference potential drop |
| $x$ | Horizontal coordinate in the deformed image |
| $X$ | Horizontal coordinate in the reference image |
| $X_0$ | Initial horizontal coordinate of the centre of a subset |
| $U$ | Horizontal component of the rigid body motion |
| $U'$ | Horizontal component of the discontinuity vector |
| $y$ | Vertical coordinate in the deformed image |
| $Y$ | Vertical coordinate in the reference image |
| $Y_0$ | Initial horizontal coordinate of the centre of a subset |
| $V$ | Vertical component of the rigid body motion |
| $V'$ | Vertical component of the discontinuity vector |
| $r^*$ | Minimum distance from the discontinuity to the subset centre |
| $\theta^*$ | Angle of the discontinuity inside the subset |
| $\varepsilon_{op}$ | Opening strain |
| $\varepsilon_{cl}$ | Closure strain |
| $R$ | Electrical resistance |
| $\rho$ | Electrical resistivity |
| $L$ | Specimen length |
| $S_{uncracked}$ | Uncracked area |
| $U_{op}$ | Closure ratio defined by the opening strain |
| $U_{cl}$ | Closure ratio defined by the closure strain |
| $\varepsilon_{max}$ | Maximal strain |
| $\varepsilon_{min}$ | Minimal strain |
| $\Delta\varepsilon_p$ | Plastic strain variation |
| $\sigma_{op}^*$ | Equivalent stabilised opening stress |
| $\sigma_{max}^*$ | Equivalent stabilised maximal stress |
| $\sigma_0$ | Flow stress |
| $\sigma_{op}$ | Opening stress |
| $\sigma_{max}$ | Maximal stress |
| $R$ | Load ratio |

**Abstract:**

The crack closure effect of a low-alloyed steel subjected to low-cycle fatigue loading has been characterized at two different imposed strain amplitudes. Two techniques (non-continuous DIC H-DIC and DCPD) have been employed in this aim, leading to similar conclusions. Thus, it is shown that the crack does not remain completely closed during a part of the compressive portion of the fatigue cycle for both applied loadings. The crack opening strains, combined with the cyclic stress-strain curve of the material allowed to determine an equivalent cyclic opening stress. This confirmed that the crack opening stresses decrease with the applied maximum stress when subjected to tension-compression loading, as commonly found in the literature.

## Introduction:

Due to pressure and temperature variations, the design of major components of primary circuits in nuclear power plants requires to evaluate the eventuality of various failure modes, including fatigue. The temperature variations, such as those endured during a power plant shutdown or restart, can induce large-scale yielding, tension-compression loading. The fracture mechanics assessments of the components impose to determine first the fatigue crack propagation over the lifespan of the component and then compare that "end-of-life" defect to a conventional-sized defect (French approach) or to the critical defect size for fast fracture or instable ductile tearing depending on the component and loadings (English approach). In this light, the determination of accurate fatigue crack propagation kinetics, representative of the potential effects of negative load ratio, are then of key importance for the manufacturer and the operator in order to safely and accurately assess the fatigue life of components and therefore ensure the component integrity against the risks of fatigue, brittle fracture and unstable ductile tearing. This in particular raises the question of the magnitude of crack closure effects under such loading conditions. Since the pioneering work of Elber [1], while crack closure has been more extensively studied under small scale yielding condition and by considering positive load ratio [1], [2], [3], [4], [5], the role of this crack-tip shielding mechanism has also been investigated for tension-compression loadings, both experimentally [6], [7] and numerically [8]. It was shown [6] that the crack may remain open for remote applied compressive loading. The crack opening load decreases with the load ratio [7]and applied loading [7], [8], [9].

The results of these studies somehow contradict the assumption generally made that the compressive part of the fatigue cycle does not contribute to the crack propagation and can then be ignored when determining the crack growth driving force parameter [10]. Although the authors reached to the same conclusions, different methods for determining the opening and closing of the crack can be implemented. Indeed, the crack opening loads were determined from the compliance variation in [6], as measured from strain gauges glued on the surface of the specimen, near the crack plane, while Paluskiewicz et al. [7] used an extensometer to measure the CMOD and derive the compliance variation. Vormwald and Seeger [9] also used small strain gauges glued on the gauge section of uniaxial specimens close to short fatigue cracks and compared the local force-strain hysteresis loops with the global loops. More recently, the Digital Image Correlation (DIC) was used [11] to characterize the local displacements and strains on a larger area on the specimen gauge section and closer to the crack tip. This technique has been used to determine the strain maps around the crack tip and quantify the plastic wake, e.g. the residual displacements created by the progressive crack propagation through the successive cyclic plastic zones [12], [13], and to measure the CMOD (crack mouth opening displacements) along the crack path [14], [15], [16], [17]and then derive the crack opening loads close to the crack tip [16], [17], [18], [19], [20] . However, different other methods, such as acoustic emission [21], [22] or DCPD can be employed. Although more generally used for the crack growth monitoring [23], [24], [25], the latter method has been successfully applied to crack closure detection [26], [27], [28], [29]and compared to other detection methods.

This study was undertaken in an attempt to characterize the crack closure behaviour of a pressure vessel low alloy steel used to manufacture steam generator shells, and to evaluate and compare two different methods for assessing crack closure. In this aim a non-continuous DIC method was first used to determine the local displacement field around the crack tip under large-scale yielding tension-compression loadings along with crack closure measurements using the DCPD technique. Based on the closure strains, an equivalent crack opening stress was then calculated and compared to various experimental results obtained for various steel and Aluminium grades.

**Material and Experimental Methods:**

*Material:* The studied material is a low-alloy18MND5 steel (equivalent to ASTM A508 Cl. 3), received in the form of a coupon extracted from a steam generator nozzle shell. The Table 1 and Table 2 provide the chemical composition as well as the monotonic tensile properties, as well as the required values from RCC-M [30].

| Element | C | S | Al | Cr | Cu | Mn | Mo | Ni | P | Si | V |
|---|---|---|---|---|---|---|---|---|---|---|---|
| Measured (wt. %) | 0.18 | <0.001 | 0.02 | 0.16 | 0.04 | 1.55 | 0.5 | 0.71 | 0.007 | 0.21 | <0.02 |
| RCC-M requirement [30] (wt. %) | ≤0.2 | ≤0.005 | ≤0.04 | ≤0.25 | ≤0.12 | 1.15-1.60 | 0.43-0.57 | 0.50-0.80 | ≤0.008 | 0.10-0.30 | ≤0.03 |

*Table 1: Chemical composition of the 18MND5 steel.*

The 18MND5 steel presents a microstructure composed of ferrite laths and lamellar cementite precipitates, gathered as bainitic packets. MnS inclusions can be observed in the material. The average prior austenitic grain size is 20 µm [31].

| | $\sigma_{YS}$ [MPa] | $\sigma_u$ [MPa] | A [%] |
|---|---|---|---|
| Measurement | 505 | 640 | 24 |
| RCC-M [30] requirement | 420 | 580 | 20 |

*Table 2 Monotonic properties of the 18MND5 low alloy steel.*

*Specimen geometry:* The specimens are first turned from cylindrical blocks prepared by wire Electrical Discharge Machining (EDM) from the shell coupons. The longitudinal axis of the cylindrical blocks is coincident with the longitudinal axis of the steam generator shell. The rectangular section is then finalized by wire EDM, and the final dimensions are shown on Figure 1. The choice of a rectangular section specimen has been governed by the use of a DIC technique. Indeed, a finer adjustment of camera optics can be achieved on the plane surface of a rectangular section specimens as compared to cylindrical specimens, where a greater depth of field is required. The specimens present a 4 mm×8 mm section, and a 16 mm gauge length. A semi-spherical notch with a depth $a_{def}$ of about 300 µm is introduced using Hole Drilling EDM in the middle of the gauge section surface and at the centre of the width. The main purposes of the notch are to facilitate the crack initiation at a chosen location, to reduce the duration of the crack initiation phase compared to a smooth specimen and, right after this initiation stage, to quickly propagate a semi-elliptic crack front.

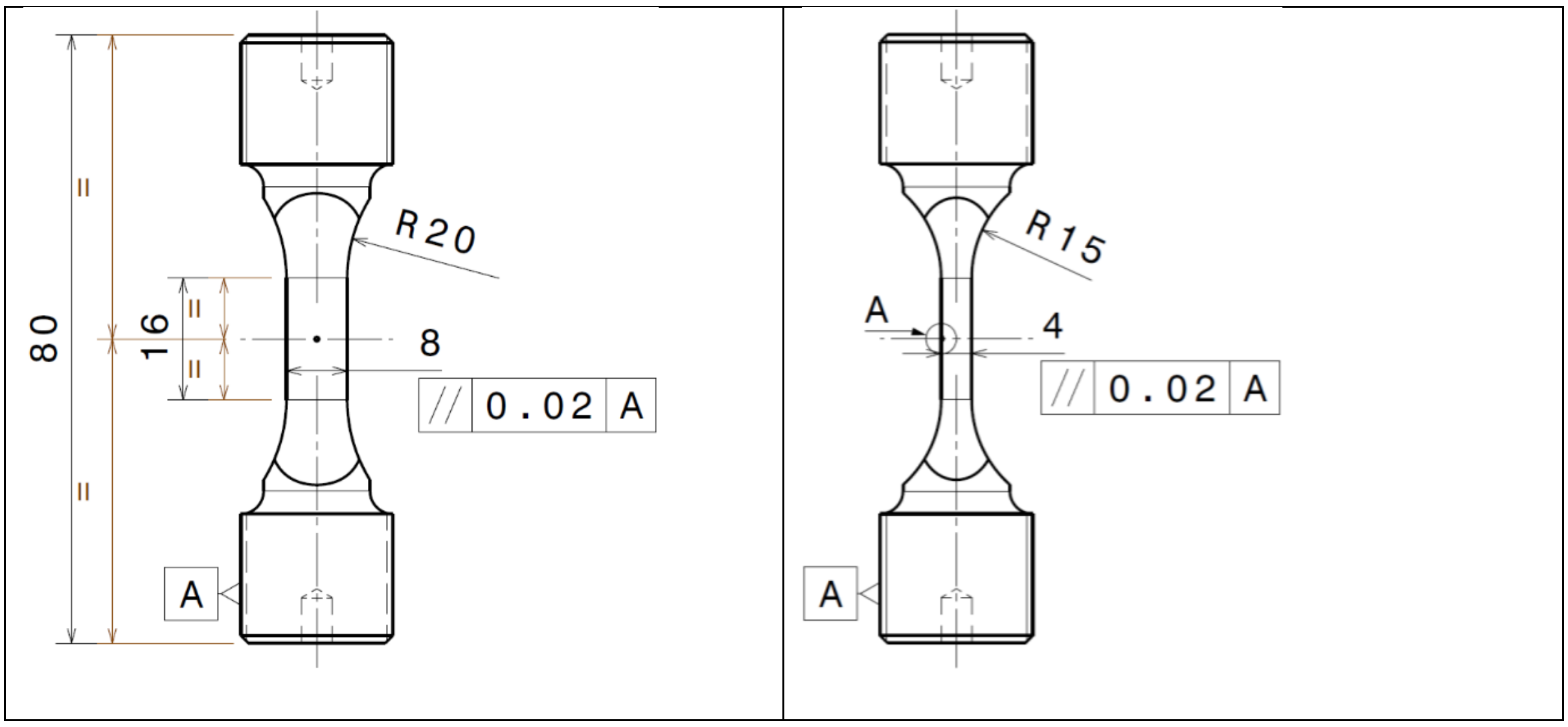

*Figure 1: Specimen geometry. All dimensions are given in unit mm.*

<u>*Experimental setup and fatigue tests:*</u> The fatigue tests were conducted in ambient air on a MTS 810 hydraulic testing machine, with a capacity of ±50 kN dynamic range. The total strain in the gauge length is controlled using an INSTRON 2620-201 extensometer. Two strain amplitudes are imposed on the gauge section of the specimens: $\Delta\varepsilon_t/2$ = 0.2 % and $\Delta\varepsilon_t/2$ = 0.6 %, with a fixed strain rate of $\dot{\varepsilon}$ = 0.4 %´s$^{-1}$. The tests were interrupted when the crack depth reached 2.6mm at the centre of the specimen. Due to the use of a DIC technique, a special attention has been paid to the test procedure. Indeed, in order to ensure a high quality of the digital images, it is necessary that the cracked section is stable during the image acquisition. As such, a specific experimental method is developed. The image acquisition is directly triggered by the software based on an electric signal delivered by the testing machine. At regular intervals, defined by a 2 % increase in the potential drop, a special cycle is introduced in the baseline loading. This cycle is divided in 41 steps, as described on the Figure 2, regardless of the applied strain amplitude. Successive static strain steps of a 2 s duration are executed, so as to acquire a static image of the vicinity of the crack at the surface of the specimen. Between every step, a strain ramp is executed at the nominal strain rate.

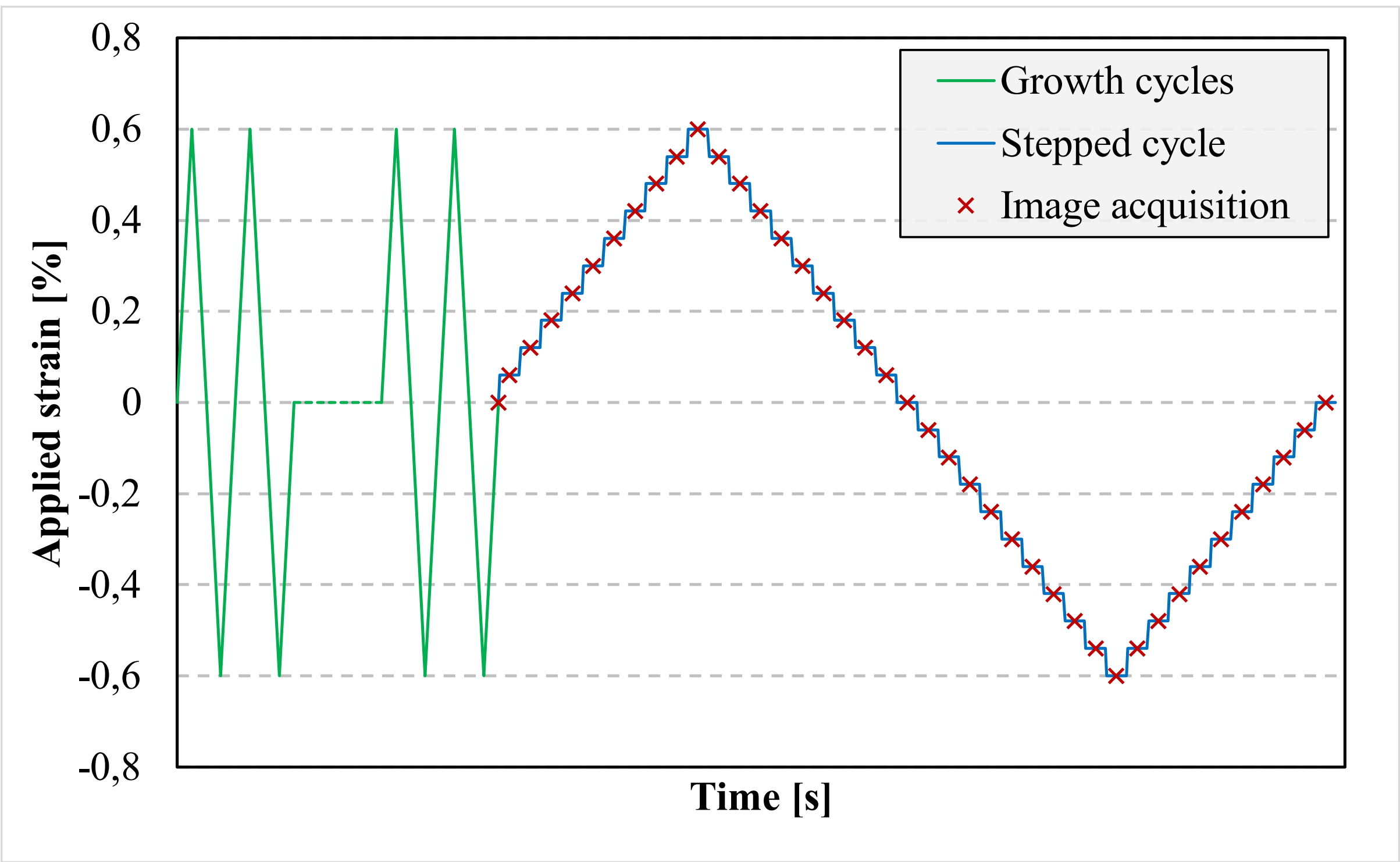

*Figure 2: Loading pattern.*

The crack size is monitored by the DCPD technique. Two 100 µm Pt wires are spot-welded on the surface of the specimen, each at a distance of 1000 µm from the centre of the notch (Figure 3a). This method furthermore requires a prior calibration of the relation between crack size and potential, which was executed using a coupled experimental-numerical approach. The numerical calibration allowed to determine a continuous calibration curve, which is validated punctually by the experimental measurements. The experimental calibration of the DCPD method has been performed using acrylic ink markings of the cracked areas, based on the protocol proposed by Cussac et al. [25]. Acrylic ink is thus brushed on the gauge section of the specimen during the test, allowing the ink to quickly spread and dry on the crack lips. At the end of the fatigue test, the specimens are immerged in liquid Nitrogen (-196 °C) and broken by applying a lever arm. The crack front markings produced using this technique are used to precisely correlate the measured crack depth and the potential drop value. Two or three markings can be realized on a same specimen, each corresponding to a point of the calibration curve. From these markings, the evolution of the crack shape can be characterized. As the observed crack fronts present a semi-elliptic shape (an example is shown in Figure 3b), a finite element model of a half specimen gauge section containing a semi-elliptic crack was developed. In this analysis, the initial crack shape is assumed semi-circular. The computations were performed using the Cast3M® software [32], using the thermoelectric analogy. A constant current (thermal flux) is applied on the top end of the gauge section, while the bottom end is electrically grounded. The potential is measured at 1 mm on each side of the crack plane, in the vertical symmetry plane of the specimen, consistently with the experimental setup.

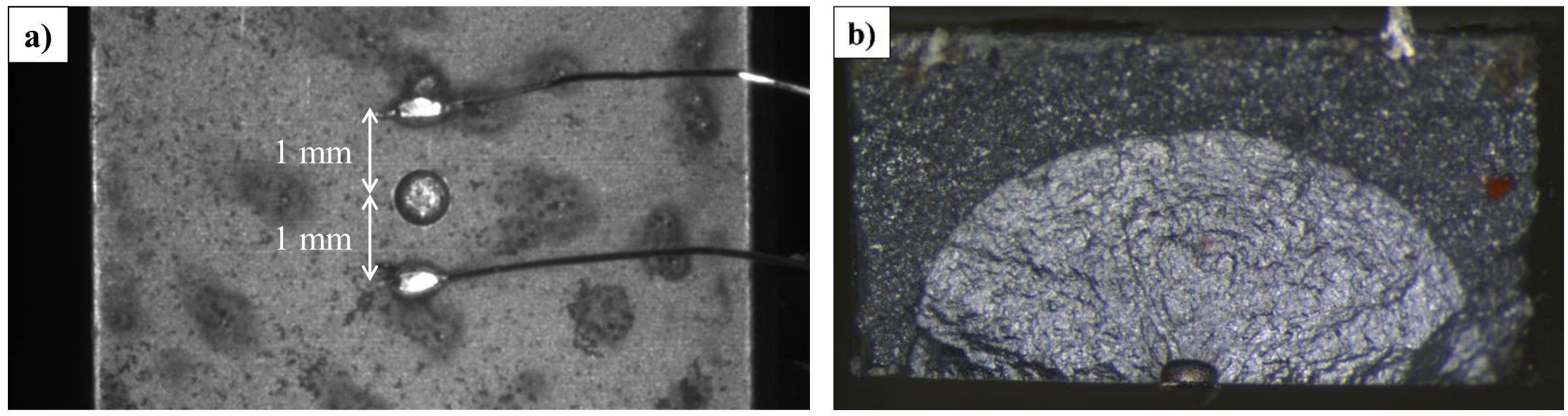

*Figure 3: a) DCPD wires spot-welded around the notch, b) Fracture surface of 18MND5 sample submitted to $\Delta\varepsilon_t/2$ = 0.6 %.*

The potential drop between the crack lips is then determined and thereafter normalised by the reference potential determined on the uncracked specimen model. Two numerical calibration curves are compared to the experimental curve, one obtained with the hypothesis of a perfectly circular crack front, and the other one with the hypothesis of a semi-elliptic crack front. In the latter case, the applied crack shapes are taken from experimental data. For the largest measured crack depth, the associated ellipticity ratio $c/a_{tot}$ is 1.18. A good agreement is observed, as shown on Figure 4 which presents the evolution of the crack depth measured from the deepest point of the notch $a_{tot} - a_{def}$ as a function of $V/V_0$, where $a_{tot}$ is the total crack depth measured from the surface of the specimen, $a_{def}$ is the depth of the notch, $V$ is the measured potential drop and $V_0$ is the reference value of the potential drop. The residual discrepancy comes from the initial hypotheses of a) a perfectly shaped crack front; b) a perfect planarity of the crack lips and c) the variation of the electrical resistivity with total and plastic strain.

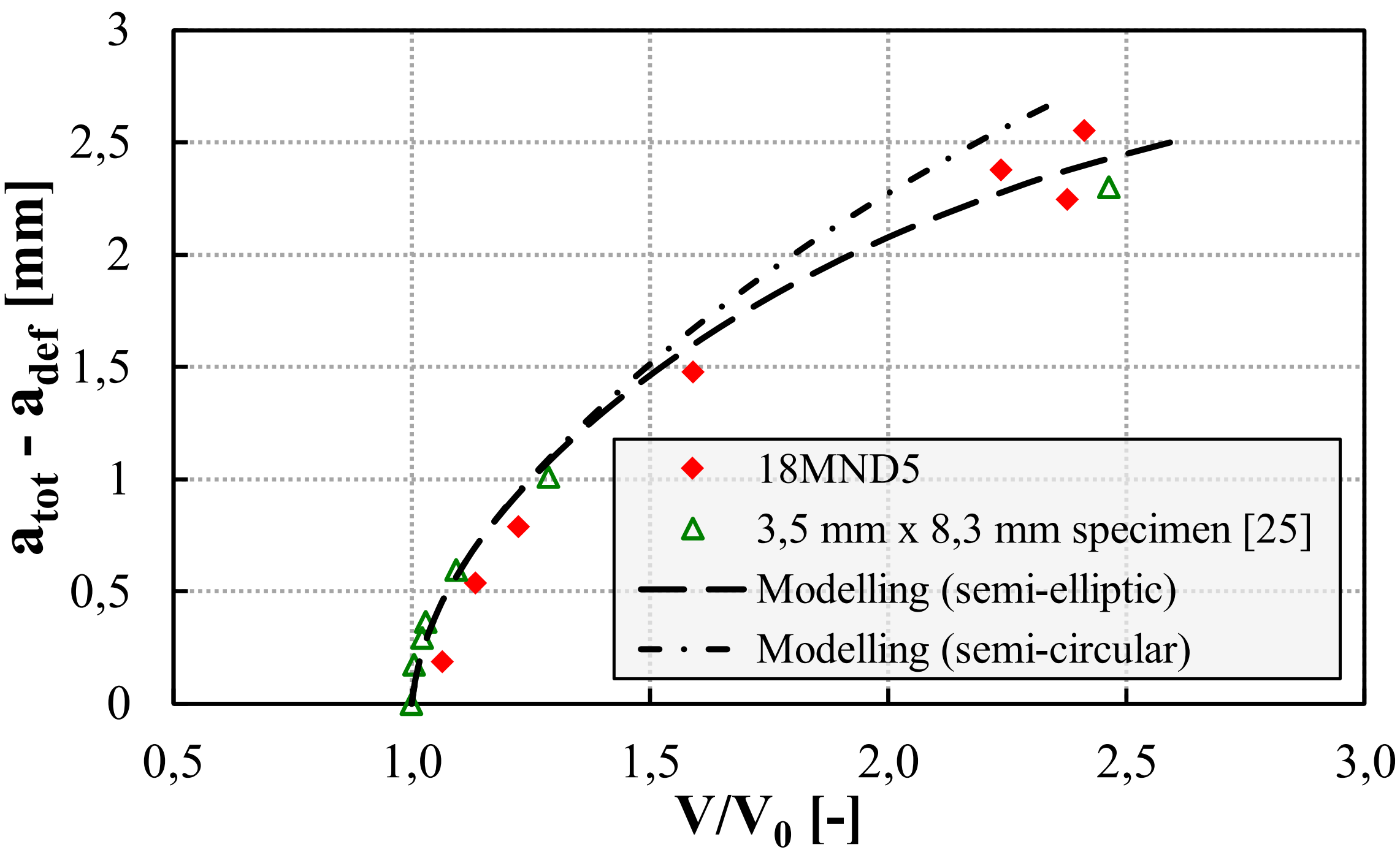

*Figure 4: Experimental and numerical calibration curves from rectangular specimen.*

*Crack opening/closure detection:* Crack opening/closure measurements were carried out using two different methods. First, a Heaviside-DIC [33] (H-DIC) method was used to extract the displacements on the surface of the specimen as this method bypasses most of the difficulties encountered by using conventional DIC. Indeed, the shape functions of the image correlation assume a continuity in the displacement field which is not verified in the case of cracked structures. As a consequence, the post-processing of conventional DIC data to derive crack opening values is strongly dependent on the obtention of a precise displacement field. The post-processing of continuous DIC results in presence of a crack has been extensively studied [34], and often relies on sophisticated techniques such as thresholding the displacement gradient standard deviation along multiple potential crack paths [35]. The Heaviside-DIC method overcomes the difficulties caused by the continuous displacement hypothesis of conventional DIC by modifying the shape function. An additional term, containing a Heaviside function, is added to the standard shape function (highlighted in blue in Eq 1), where only the first order of the gradient of displacements is considered.

$$\begin{cases} x = X + U + \frac{\partial U}{\partial X}(X - X_0) + \frac{\partial U}{\partial Y}(Y - Y_0) + U' \cdot H^1((X - X_0), (Y - Y_0)) \\ y = Y + V + \frac{\partial V}{\partial X}(X - X_0) + \frac{\partial V}{\partial Y}(Y - Y_0) + V' \cdot H^1((X - X_0), (Y - Y_0)) \end{cases} \quad (1)$$

Where $X, Y$ are the coordinates in the reference image, $x, y$ are the coordinates in the deformed image and $X_0$ and $Y_0$ are the initial coordinates of the centre of the subset. $U$ and $V$ are the horizontal and vertical components of the rigid body motion. $U'$ and $V'$ are respectively the horizontal and vertical components of the discontinuity vector describing the jump in the displacement field. $H^1$ is a two-dimensional Heaviside function, defined in Eq. 2:

$$H^1\big((X - X_0), (Y - Y_0)\big) = H^1((X - X_0) \cdot \cos(\theta^*) + (Y - Y_0) \cdot \sin(\theta^*) - r^*) = H^1(r - r^*)$$

$$= \begin{cases} 1, & r \leq r^* \\ 0, & r > r^* \end{cases} \quad (2)$$

The location of the discontinuity is described in the subset by its distance to the centre $r^*$ and the angle $\theta^*$as shown in Figure 5.

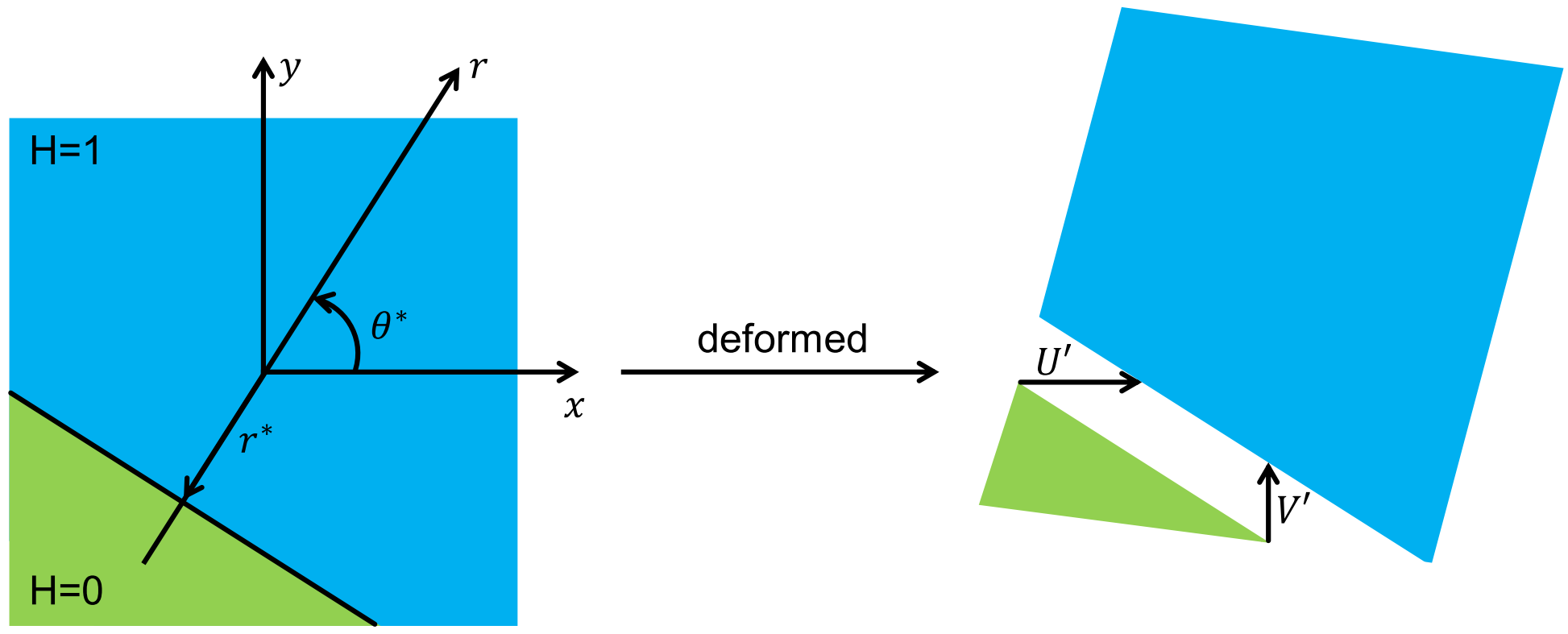

*Figure 5: Heaviside-DIC principle (from [33] and [36]). U' and V' are respectively the horizontal and vertical components of the discontinuity vector.*

The crack opening displacements $U'$ and $V'$ are therefore a direct result of the H-DIC correlation calculation and can be exploited to determine the crack tip and the crack lips position. The EDM machining of the specimens generates a natural speckle on the surface that is sufficient for the analysis software to track the displacements in the zone of interest. A low noise is obtained, as is shown on Figure 6, showing the crack opening values map in the loading direction in a zone centered around the crack tip. A clear distinction between the crack and the bulk material is then observed, so that the crack lips position can be accurately identified by thresholding the crack opening values at a low level.

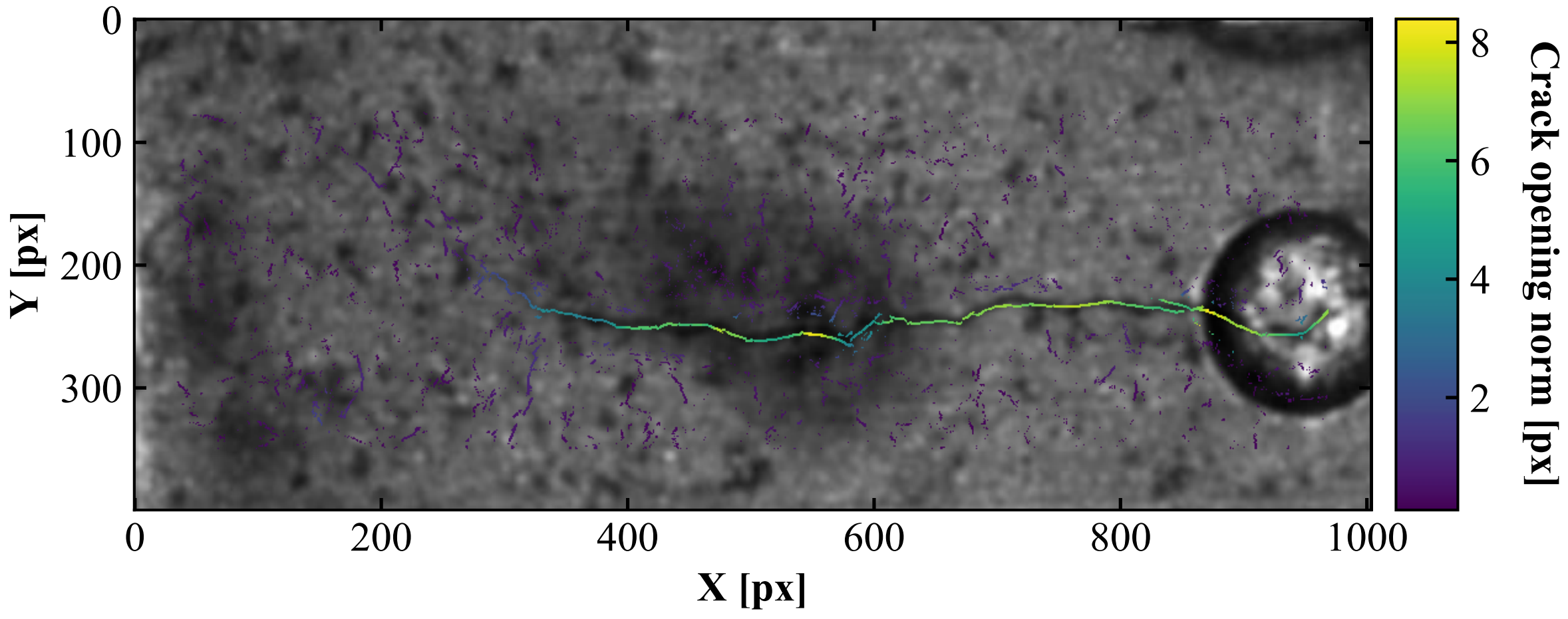

*Figure 6: Evolution of the crack opening along the crack front.*

The determination of the crack opening maps, coupled with the acquisition of 41 images per cycle, finally leads to the obtention of the imposed strain versus crack opening at regular propagation intervals. The crack opening is extracted at 100 µm from the crack tip. Due to the reduced crack opening displacement of small cracks, the minimal crack depth required for a proper analysis of strain-opening hysteresis loops is of 1 mm. In the case of a shallower crack, the crack opening presents a large noise, making the data processing too complex. As an example, a strain-CMOD loop obtained during a strain-controlled test at $\Delta\varepsilon_t/2$ = 0.6 % for a 1.65 mm deep crack is presented in Figure 7a). Two distinct domains can be identified:

- The first one where the crack opening displacement remains stable near 0 µm;
- the other one, during which the crack progressively opens and closes.

The opening and closure strains $\varepsilon_{op}$ and $\varepsilon_{cl}$ are in this case equal to -0.42 %. The crack then closes while the gauge length of the specimen is still subjected to a nominal compression, as shown by the opening and closure points on the corresponding stress-strain hysteresis loop (Figure 7b), highlighted in rose and blue, respectively.

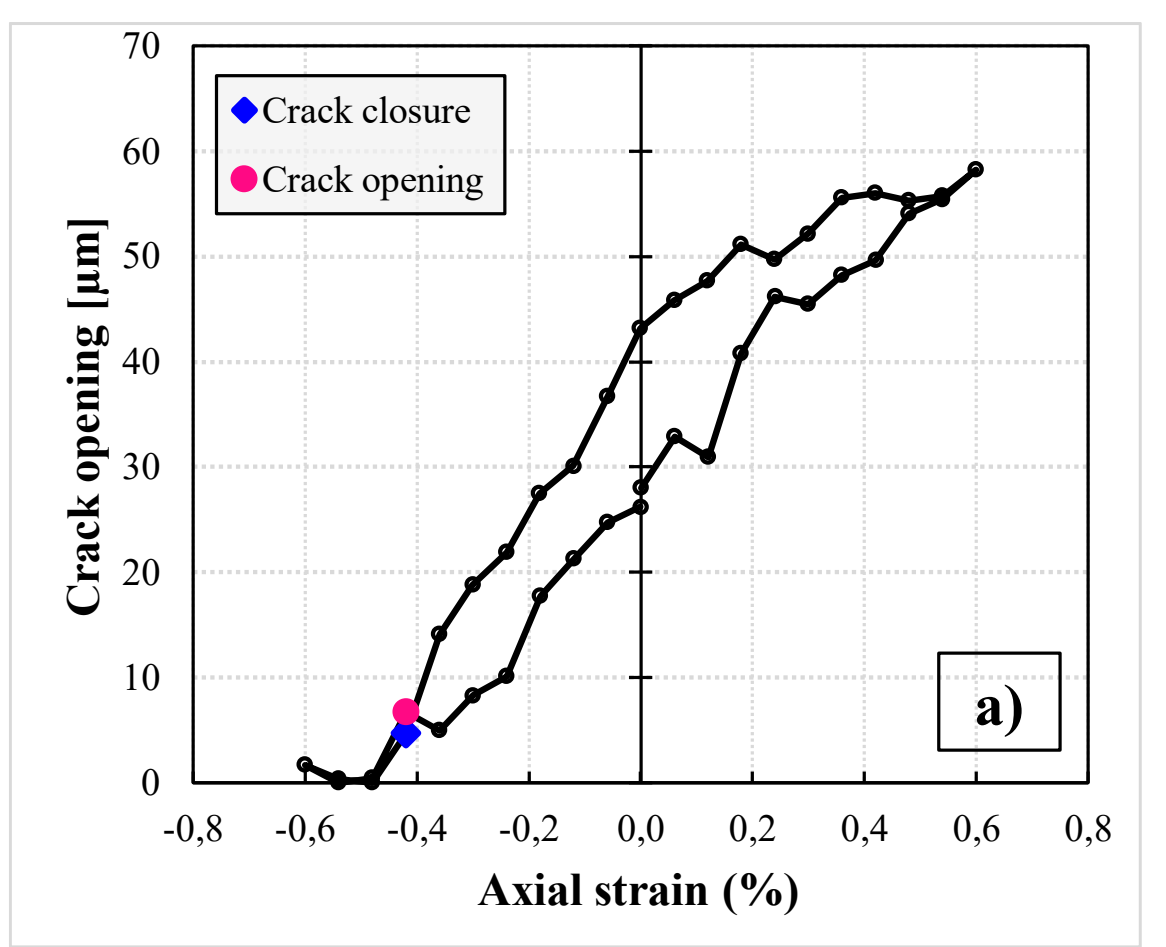

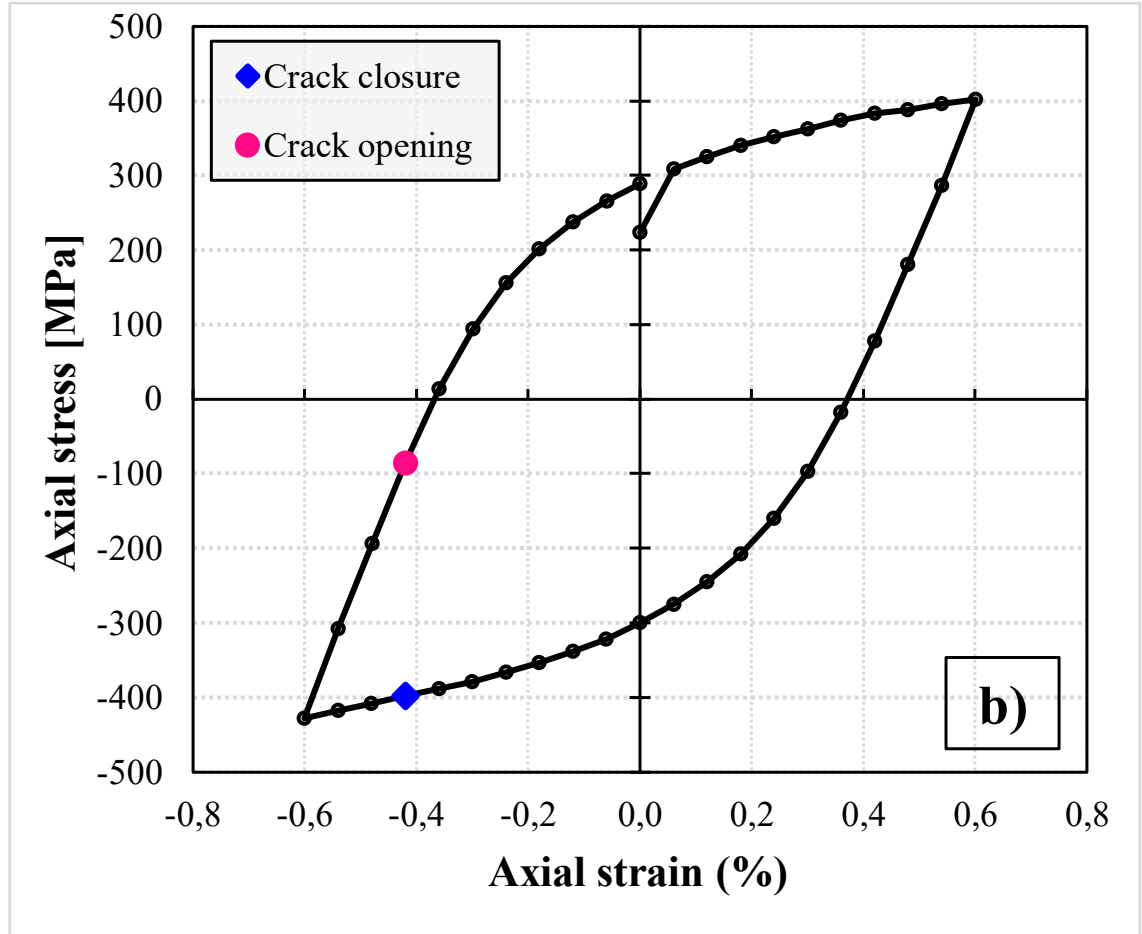

*Figure 7: a) Crack closure detection for a 1.65 mm deep crack at $\Delta\varepsilon_t/2$ = 0.6 % and b) view of opening and closure points on the corresponding macroscopic stress-strain hysteresis loop.*

In addition, the DCPD signal recorded during the constant strain phases of the tests (imaging phases) were processed. The evolution of the DCPD with the applied strain is plotted in Figure 8, and is consistent with the one observed by Andersson et al. in a IN718 alloy at elevated temperature [37]. The first rise of the DCPD obtained for strain values ranging between -0.6 % and -0.48 % is caused by the closure of the crack. Indeed, as the crack opens, the contact surface decreases, and therefore the potential drop increases since the resistance of the specimen is inversely related to the contact area, as shows Eq. 3.

$$R = \frac{\rho L}{S_{uncracked}} \quad (3)$$

Where: $\rho$ is the resistivity of the material, L the length of the specimen and $S_{uncracked}$ the uncracked area.

The second linear evolution of the DCPD with the total applied strain is caused by the variation of the resistivity due to the increase of the total strain [38], [39]. A threshold in terms of minimum crack depth for opening detection, depending on the applied loading, is however observed. Indeed, for $\Delta\varepsilon_t/2$ = 0.2 % the threshold is $a_{tot} \geq 2$ mm, while for $\Delta\varepsilon_t/2$ = 0.6 % the threshold value is $a_{tot} \geq 1.4$ mm. For lower crack depths, the evolution of the DCPD signal with the applied strain is purely linear and presents no slope variation. The evolution of the DCPD signal is then only caused by the variation of the electrical resistivity due to the increase of the total strain.

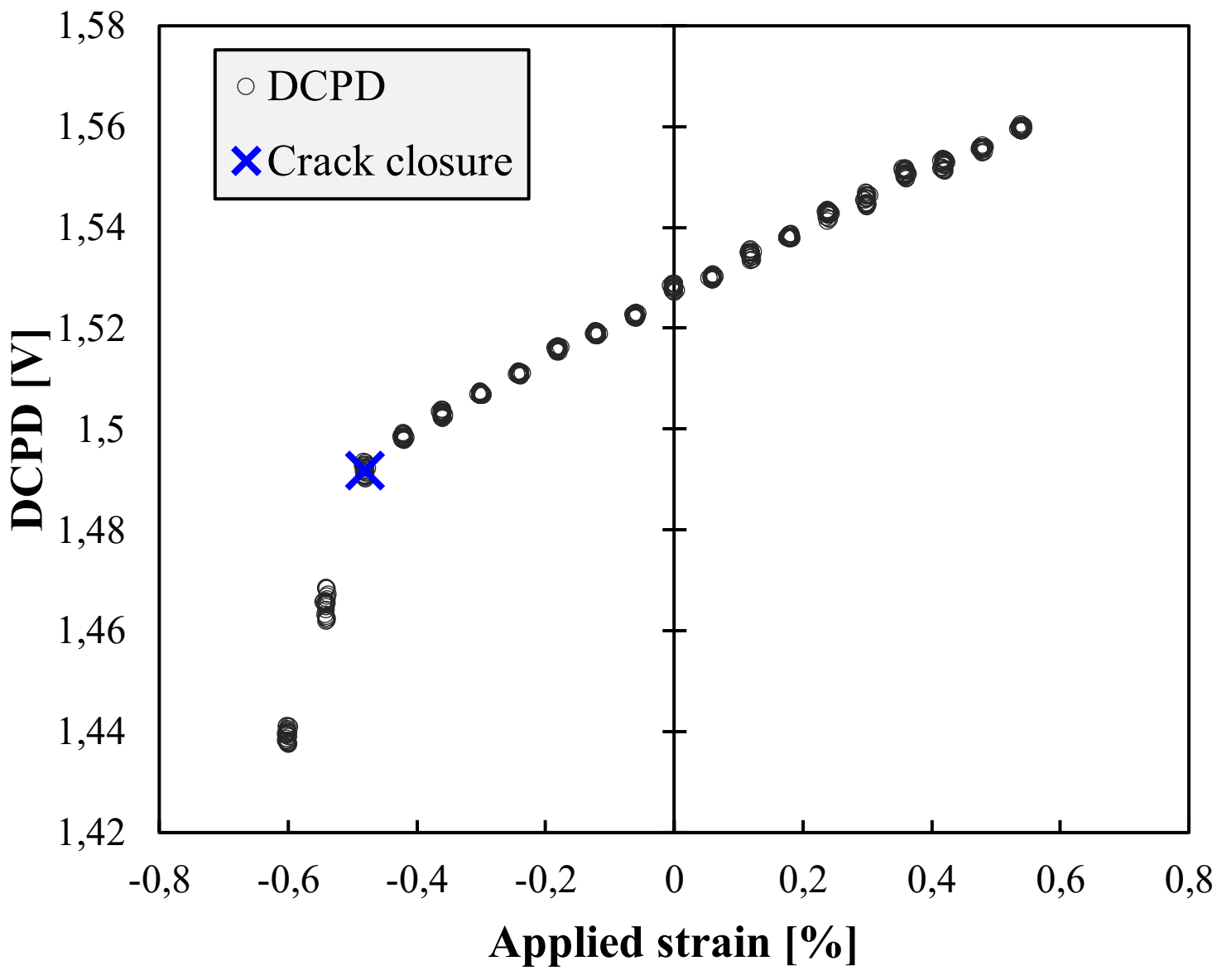

*Figure 8: Evolution of the DCPD signal with the applied strain for a 1.65 mm deep crack.*

**Results:**

Two fatigue tests were conducted at total strain amplitudes of $\Delta\varepsilon_t/2$=0.2 % and $\Delta\varepsilon_t/2$=0.6 %. The experimental setup described above is employed, and crack opening and closure strains are extracted at regular crack propagation intervals using both H-DIC and DCPD methods. It was first noticed that the H-DIC method allows to determine the crack opening and closure for smaller cracks than by using the DCPD method. The evolution of the opening and closure strains are plotted as a function of the total crack depth measured from the specimen surface $a_{tot}$ in the Figure 9a) for $\Delta\varepsilon_t/2$ = 0.6 % and Figure 9b) for $\Delta\varepsilon_t/2$ = 0.2 %. Despite a noticeable scatter, it appears that the crack opening and closure strains decrease as the crack grows in depth, indicating that the crack remains opened during a larger portion of the cycle as the crack propagates. The scatter comes from the use here of a natural speckle, whose size may be small in comparison with respect to the displacement values to be measured. Indeed, the grey level distribution (ranging from 0 to 255) presents a standard deviation of around 22, meaning that a generally low contrast is obtained in the studied images. An indicator of the correlation error is defined as the standard deviation of the crack opening displacement over the whole image. Despite this relatively low contrast, the standard deviation of the crack opening displacement is of 0.3 pixels, or around 2 µm in the present case. For comparison, the standard deviation of the crack opening obtained for an ideal simulated case in the reference study for the XCorrel software [33] is 0.1 pixel, while [36] obtained a standard deviation value of 0.4 pixel. The use of an improved speckle, with a larger size, might result in a reduced scatter. It is shown that the values of the opening and closure strains are of the same order of magnitude and present the same trends, for both strain amplitudes. In particular, the crack opening and closure strains tend to be negative, suggesting that a part of the compressive part of the cycle must be taken into account when calculating a parameter representative of the crack driving force.

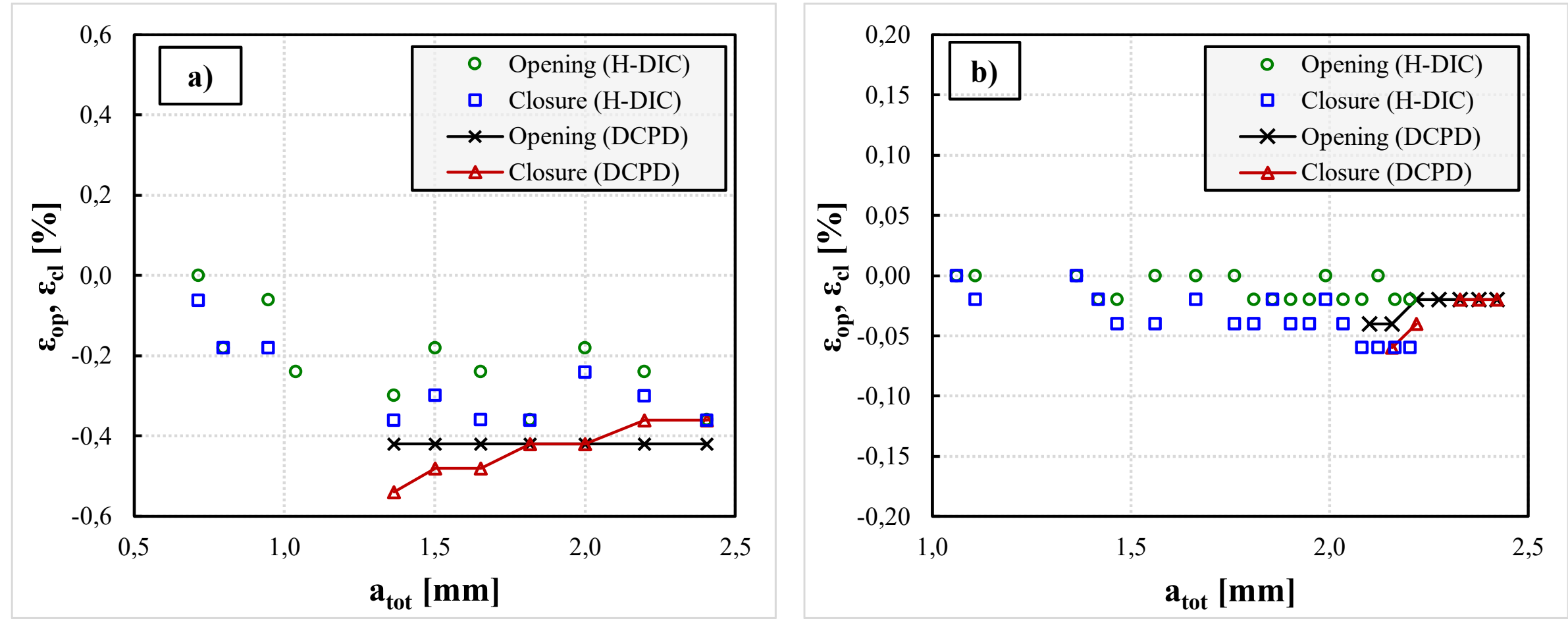

*Figure 9: Crack opening and closure strains measured by H-DIC and DCPD at a) $\Delta\varepsilon_t/2$ = 0.6 % and b) $\Delta\varepsilon_t/2$ = 0.2 %.*

The evolution of the opening and closure strains shown in Figure 9 are consistent with the trend observed by Minakawa et al. [40], where the crack opening stresses or strains decrease as the crack propagates.

Two values of crack opening ratios are thereafter considered, based on the opening and closure strains, and are calculated according to Eq. 4.

$$U_{op} = \frac{\varepsilon_{max} - \varepsilon_{op}}{\varepsilon_{max} - \varepsilon_{min}} \text{ and } U_{cl} = \frac{\varepsilon_{max} - \varepsilon_{cl}}{\varepsilon_{max} - \varepsilon_{min}} \quad (4)$$

These quantities allow to evaluate the normalized portion of the cycle during which the crack remains opened. The variations of the $U_{op}$ and $U_{cl}$ values as a function of crack depth are plotted in Figure 10.

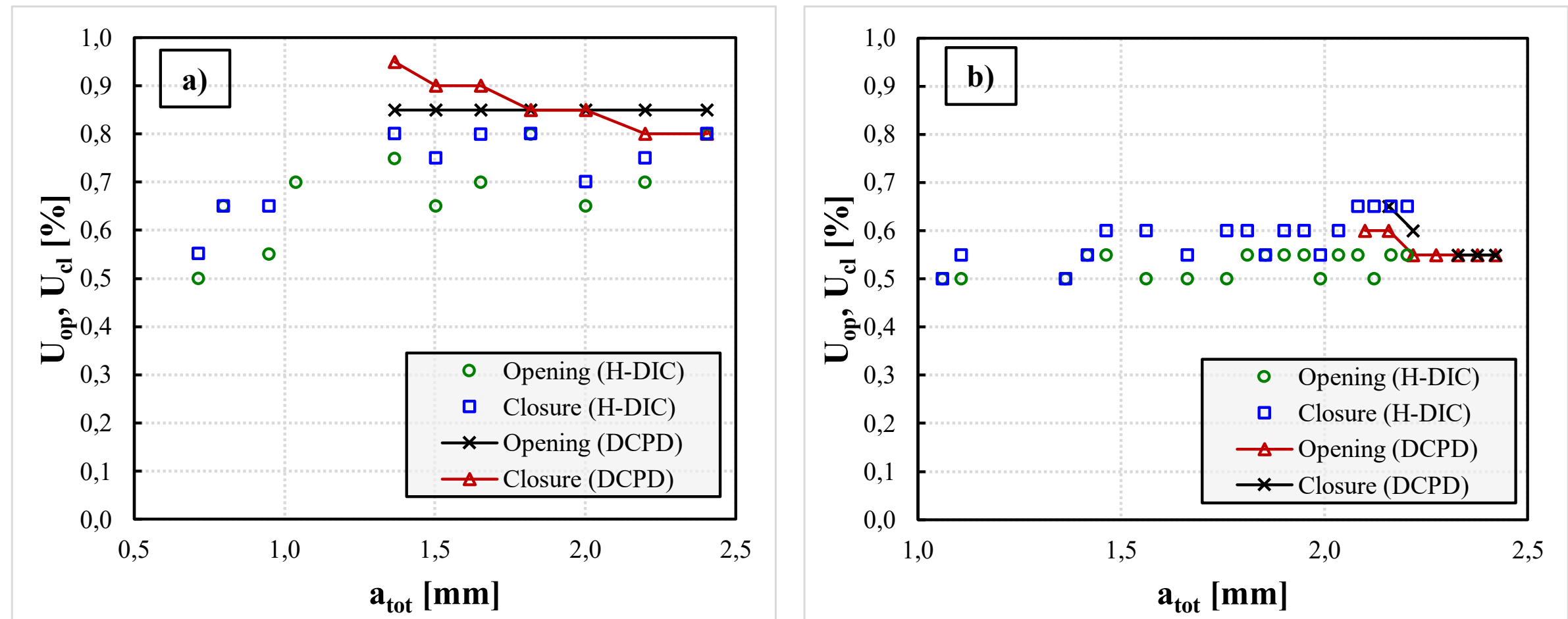

*Figure 10: Crack opening ratios $U_{op}$ and $U_{cl}$ determined using H-DIC and DCPD for a) $\Delta\varepsilon_t/2$ = 0.6 % ($\Delta\varepsilon_p/2$ = 0.30 %) and b) $\Delta\varepsilon_t/2$ = 0.2 % ($\Delta\varepsilon_p/2$ = 0.00 %).*

**Discussion:**

Two different techniques, namely H-DIC and DCPD, have been used here to determine crack opening during low-cycle fatigue loading. The results presented in the previous section indicate that the values of crack opening strains are slightly lower when derived from the DCPD variation during a loading cycle as compared to H-DIC. The measured potential is directly linked to the remaining uncracked surface; hence it provides a global measurement of the crack opening strain along the crack front while the H-DIC is a surface measurement It has been extensively shown, both by experiment and calculations [41], [42], [43], [44], that the crack opening stress or strain is dependent on the stress state. Indeed, the crack opening stresses/strains are higher at the surface of the specimen, where the stress tensor tends towards a plane stress state while being influenced by the bulk of the specimen, which is dominated by a plane strain state. As the opening strain value determined using DCPD can be considered as a global value averaged along the crack front, this value is lower than the plane stress-dominated value determined using H-DIC. Nevertheless, a good and consistent agreement is found between the H-DIC and DCPD-determined crack opening strains.

For the two applied strain amplitudes, the variation of crack opening/closure strains with the crack propagation present the same trends, although the stabilized opening/closure levels are highly dependent on the imposed strain amplitude. It is worth noticing that the higher amplitude is associated with large-scale yielding condition in the gauge section, while for the lower strain amplitude, it can be considered that small-scale yielding condition is prevailing, as indicated by the stabilized stress-strain hysteresis loops presented in Figure 12. The higher the plastic strain amplitude (respectively around 0 % and 0.30 %), the higher the crack opening ratio is. This is consistent with the findings of Watson et al. [45], who showed, for tension-compression loading in a 304 austenitic stainless steel, that the crack opening ratio increases with the applied strain amplitude and the plastic strain amplitude. For total strain-controlled tests, the crack opening is best described in terms of strain and not stress [9], [19]. In this light, an equivalent stabilized crack opening stresses $\sigma_{op}^*$ and a maximum stress $\sigma_{max}^*$ are estimated from the cyclic stress-strain curve of the material shown in Figure 11, obtained during LCF testing on smooth specimens [25].

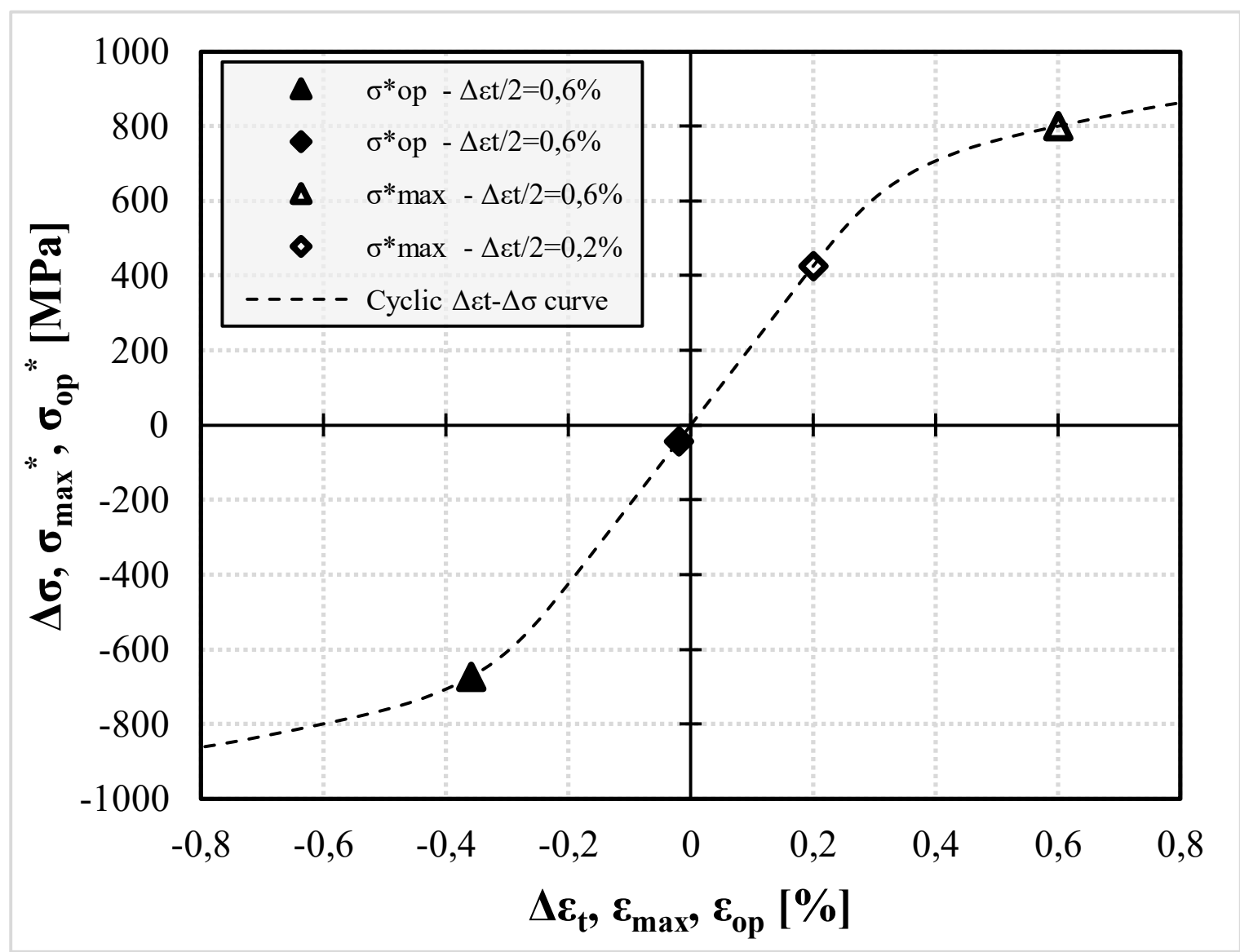

*Figure 11: Cyclic stress-strain curve of 18MND5 low-alloy steel and opening strains and equivalent stresses.*

These information can be plotted in the representation proposed by Vormwald and Seeger [9], showing the opening stress variation as a function of the maximum stress for various materials. The stresses are normalized by the flow stress $\sigma_0$ of the considered materials, which is determined according to Eq. 5.

$$\sigma_0 = \frac{1}{2} \times (\sigma_{YS} + \sigma_u) \tag{5}$$

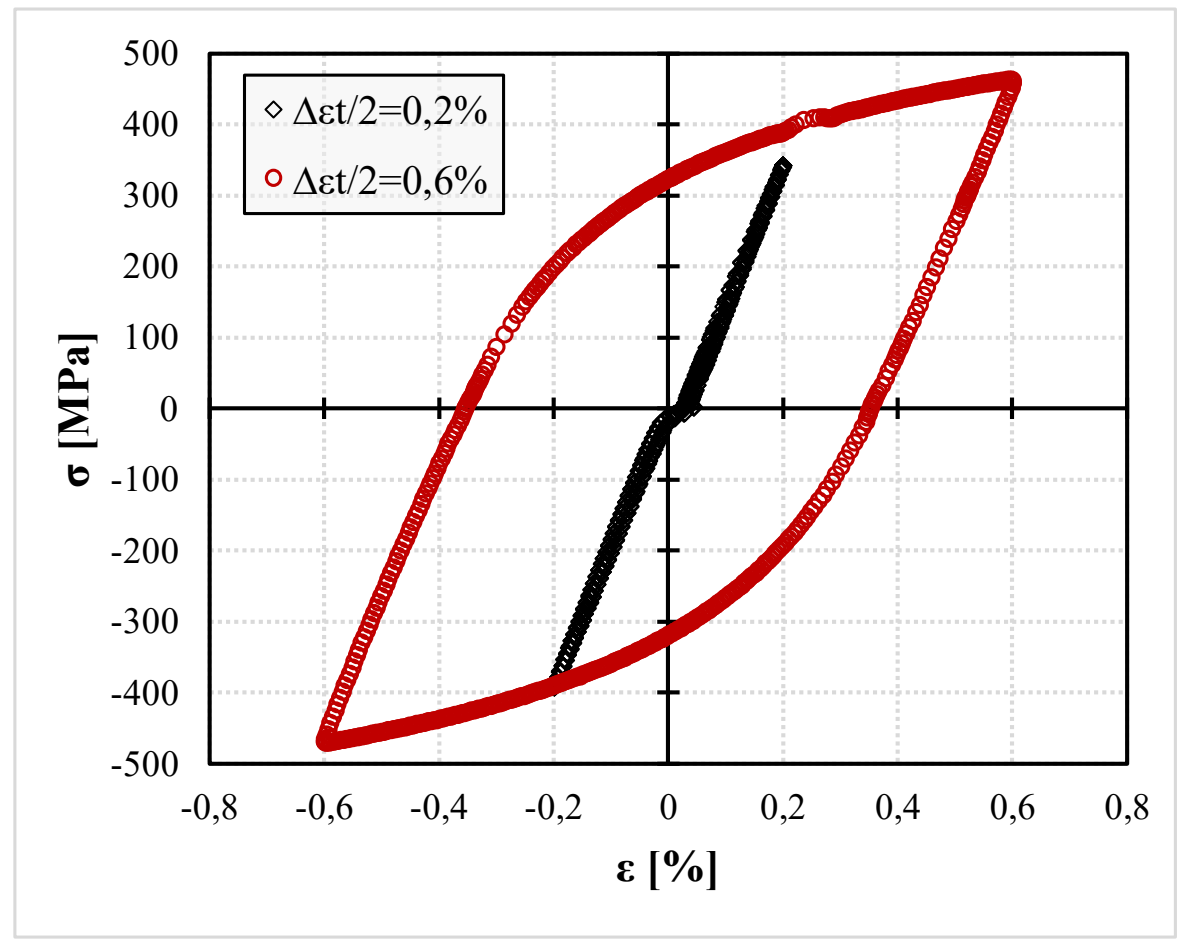

*Figure 12: Stabilized stress-strain hysteresis loops [25].*

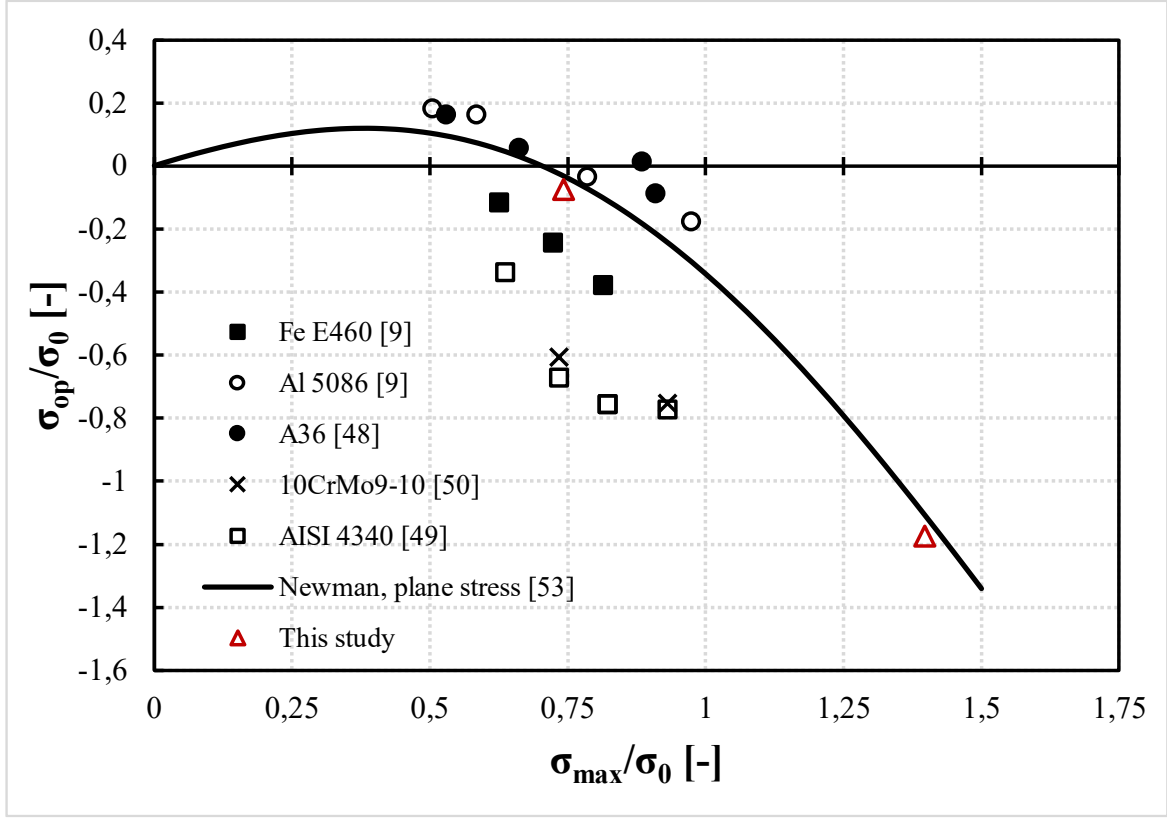

*Figure 13: Evolution of the crack opening stress as a function of the maximum applied stress for various steel grades.*

As shown in Figure 13, the evolution of the opening stress for constant-amplitude tests carried out at $R_{\varepsilon}$ = -1 is consistent with the trends observed in early studies on the topic [9], [46], [47], [48], where it was already shown that the crack opening stress decreases when the maximum applied stress tends to the flow stress of the material. Additionally, in all cases the crack opening stress becomes negative when the maximum applied stress approaches the flow stress, which was also observed in the present study. The equivalent normalized crack opening stresses determined in this study lie within the range of values obtained for steels [9], [46], [47], [48] and Al alloys [9].

Although the presented trends are the same, a discrepancy is still noted between the present study and the literature data. Two factors are identified as the causes for this discrepancy. The first one is the definition of the crack opening points. The crack opening detection methods employed varied between the studies, as for example [46] defined the crack opening as the first occurrence of positive value of the CMOD as close as possible to the crack tip, while [47] also measured the CMOD, but at 100 μm from the crack tip. The kinematic of the crack opening motion implies that the determined crack opening stresses are slightly lower in the latter case. This difference might still be limited, as the present study only shows a slight difference between crack opening strains obtained through DCPD and CMOD at 100 µm from the crack tip. In the work presented in [48], the crack opening was detected by using the ACPD and compliance methods, while [9] determined the crack opening by comparing the local σ-ε hysteresis loops (using a microstrain gauge) with the global ones (using an extensometer). A second factor is the strain-hardening behaviour that varies between the materials studied in [9], [46], [47] and [48]. Indeed, as shown by Pommier and Bompard [49], the crack opening and closure stresses of a given material are influenced by the respective contribution of isotropic and kinematic cyclic hardening. Also, the strain hardening coefficient has a significant influence on the crack opening stresses as shown by Reichenbacher et al. [50]. It is demonstrated that for a material with a higher strain hardening coefficient, the crack opening stress are lower in comparison to a material presenting a low strain hardening coefficient. This is also observed experimentally in [9], where the FeE460 alloy exhibits a lower crack opening stress than the Al5086 alloy, which presents less strain-hardening. The crack opening stress value is furthermore in agreement with the one predicted value by Newman (Eq. 6) [51] in plane stress conditions.

$$\frac{\sigma_{op}}{\sigma_{max}} = A_0 + A_1 \times R \; for -1 \leq R \leq 0 \tag{6}$$

With:

$$A_0 = 0.535 \times \cos\left(\frac{\pi\sigma_{max}}{2\sigma_0}\right) \; and \; A_1 = 0.344 \times \frac{\sigma_{max}}{\sigma_0}$$

Additional data (based on this study and extended to other materials) are nevertheless required to confirm this point.

## Conclusions:

The fatigue crack closure effect under reversed tension-compression loading has been studied in the case of a nuclear-grade low-alloy steel. The crack opening and closure strains were thus experimentally determined using the non-continuous H-DIC method, in an attempt to circumvent the difficulties caused by conventional continuous DIC methods, as well as the direct current potential drop method for two applied total strain amplitudes. Indeed, the Heaviside function implemented in the shape functions results in a direct determination of the crack opening, avoiding complex data processing methods required by the conventional (continuous) approach. The main findings can be summarized as follows:

The opening and closure strains determined using DCPD are lower than those obtained by H-DIC. This can be at least partly explained by the transition from plane stress state at the surface of the specimen to plane strain prevailing at the deepest point of the crack front which is captured by the DCPD technique, while the H-DIC measurements are mostly concerned with the plane stress region. In addition, the crack opening/closure strains decrease with the increase in imposed total strain amplitude. A method based on the cyclic stress-strain curve is used to determine an equivalent crack opening stress and compare its value with various material grades. It comes out that the evolution of the equivalent opening stress with the equivalent maximum applied stress is consistent with Newman's [51] model and experimental data from literature pertaining to steel and Aluminium grades.

The crack opening data obtained from this study will be thereafter used to determine a crack growth parameter based on J-integral accounting for the crack closure effect in order to analyse small-scale and large-scale yielding conditions.

## Acknowledgments:

This work was supported by Framatome, under the CIFRE grant n°2020/1891.

CRediT: **Théotime ASSELIN:** Conceptualization, Methodology, Investigation, Writing-Original draft preparation. **Olivier ANCELET:** Methodology, Funding Acquisition, Validation, Writing-Reviewing and Editing. **Guillaume BENOIT:** Software, Investigation, Writing-Reviewing and Editing. **Florence HAMON:** Investigation, Writing-Reviewing and Editing. **Valéry VALLE:** Software, Validation, Writing – Reviewing and Editing. **Gilbert HÉNAFF:** Methodology, Funding Acquisition, Validation, Writing-Reviewing and Editing.